# Magnetic instabilities in doped Fe$_2$YZ full-Heusler thermoelectric compounds

Sébastien Lemal,[1,][*] Fabio Ricci,[1,][*] Daniel I. Bilc,[2] Matthieu J. Verstraete,[3] and Philippe Ghosez[1]

[1]*Theoretical Materials Physics, Q-Mat, CESAM, Université de Liège, B-4000 Liège, Belgium*
[2]*Faculty of Physics, Babeş-Bolyai University, 1 Kogălniceanu, 400084 Cluj-Napoca, Romania*
[3]*Nanomat, Q-Mat, CESAM and European Theoretical Spectroscopy Facility, Université de Liège, B-4000 Liège, Belgium*
(Dated: November 18, 2019)

Thermoelectricity is a promising avenue for harvesting energy but large-scale applications are still hampered by the lack of highly-efficient low-cost materials. Recently, Fe$_2$YZ Heusler compounds were predicted theoretically to be interesting candidates with large thermoelectric power factor. Here, we show that under doping conditions compatible with thermoelectric applications, these materials are prone to an unexpected magnetic instability detrimental to their thermoelectric performance. We rationalize the physics at the origin of this instability, provide guidelines for avoiding it and discuss its impact on the thermoelectric power factor. Doing so, we also point out the shortcomings of the rigid band approximation commonly used in high-throughput theoretical searches of new thermoelectrics.

Thermoelectric (TE) modules realizing the direct conversion of wasted heat into electricity appear as very promising devices for clean energy harvesting [1]. However, concrete TE applications still remain limited to niche markets due to the lack of cheap and efficient thermoelectric compounds. The efficiency of thermoelectrics is quantified by their figure of merit $ZT = S^2\sigma T/\kappa$ involving the Seebeck coefficient ($S$), the electrical conductivity ($\sigma$), the temperature ($T$) and the thermal conductivity ($\kappa$). Attempts to optimize $ZT$ by reducing $\kappa$ already led to record values in Bi$_2$Te$_3$ ($\sim$2.4) [2] and SnSe ($\sim$2.6) based systems [3]. Further improvements now imply also boosting the power factor (PF), $S^2\sigma$, using nontrivial electronic band structure engineering. The simultaneous increase of $S$ and $\sigma$ is challenging as it requires mutually exclusive characteristics [4]: abruptly changing density of states (flat bands) and large group velocity (dispersive bands).

The fast screening of the PF of a vast palette of compounds using computational methods appears as a very useful approach in order to identify new promising TE candidates with suitable performance [5–7]. This screening typically relies on first-principles calculations of the electronic properties of pristine phases, and the use of the rigid band approximation to predict the TE properties under appropriate doping [6, 8]. Using such an approach, Bilc *et al.* [9] recently identified Fe$_2$YZ full Heusler compounds as a new class of attractive candidates with large PF. The interesting properties of Fe$_2$YZ compounds were linked to the highly-directional character of the Fe 3$d$ states, leading to "flat-and-dispersive" bands compatible with Mahan's requirements [4].

In this Letter, we study from first-principles the properties of Fe$_2$YZ compounds under explicit doping, and show that they are prone to a magnetic instability which is detrimental to their TE properties. We rationalize the origin of this instability and provide guiding rules for avoiding it. Our work confirms the interest of Fe$_2$YZ compounds for TE applications, further extending it to thermo-magnetic applications. We also demonstrate that theoretical predictions based on the rigid band approximation in the pristine phase can often be qualitatively incorrect, and should be more systematically complemented by simulations under explicit doping.

*Methods.* Density Functional Theory (DFT) simulations are performed using the CRYSTAL [10, 11] and ABINIT [12] codes. With CRYSTAL, we performed hybrid functional calculations relying on the B1 Wu-Cohen [13] (B1-WC) functional, previously used for this class of materials [9]. With ABINIT, we used the Projector Augmented Wave method, and the Generalized Gradient Approximation (GGA) exchange-correlation functional of Perdew-Burke-Ernzerhof [14] with an additional Hubbard-like $U$ correction [15]. The $U$ parameter on the transition metal $d$-orbitals (namely, Ti, Zr, Hf, V, Nb, Ta, Fe, Ru, Os) is determined self-consistently by means of linear response [16]. The two approaches benchmark each other and provide structural and electronic properties in fair agreement (Tab. I in Ref. [17]).

Most results rely on the B1-WC functional: we explore doping effects from explicit atomic substitutions (*explicit* doping) using cubic and tetragonal supercells [17] yielding average dopant concentrations between $3.8 \times 10^{20}$ and $1.2 \times 10^{21}$ cm$^{-3}$. For the Fe$_2$YZ$_{1-x}$A$_x$ (Fe$_2$YZ$_A$) compounds with $A$ = Si, P, Ge, Sb this corresponds to $x = 0, 1/48, 1/32$ and $1/16$.

The GGA+$U$ approach was used in a computer experiment interpolating the electronic band structures of the different Fe$_2$YZ compounds by artificially changing $U$ and to study Ru$_2$ZrSn and Os$_2$HfSn with and without spin-orbit coupling (SOC) correction. It was also used to scan a continuous range of doping concentrations by adding fractions of extra electron compensated by a positive background to the unit cell of the pristine phases. Such an *implicit* doping method bypasses the need for explicit atomic impurities and related structural distortions, so more directly probing purely electronic effects. A more detailed study is given in Ref. [17].

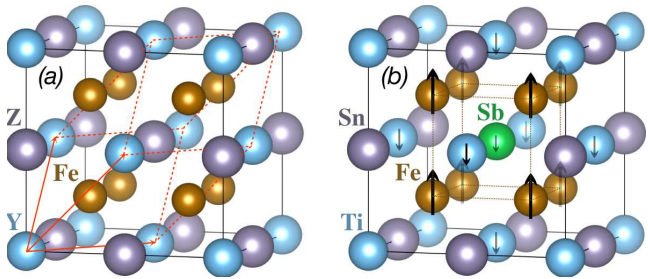

FIG. 1. (Color online) *(a)* $L2_1$ crystal structure of Fe$_2YZ$ compounds; red (black) line highlights the primitive (conventional) $FCC$ cell. *(b)* Schematic arrangement of Fe$_2$TiSn$_{Sb}$ magnetic moments.

The transport properties are computed semi-classically in the rigid band and constant relaxation time ($\tau = 34$ fs, see Supplementary of Ref. [18]) approximations with the BoltzTraP code [19].

*Concentration effects.* Consequences of explicit doping are shown in Fig. 2 reporting the Fe$_2$TiSn$_{Sb}$ density of states (DOS) at different concentrations. The pristine phase *(a)* is semiconducting and non-magnetic (NM) (it obeys the Slater-Pauling rule [20, 21]) with a band gap of 1.04 eV between Fe $t_{2g}$ and Fe $e_g$ states at the valence band maximum (VBM) and conduction band minimum (CBM) respectively [22–24]. In order to perturb as little as possible the band edge states near the Fermi level ($E_F$) responsible for the high PFs [9], we choose to dope it by partly substituting Sn with Sb on the $Z$-site.

In a NM calculation (Fig. 2*(b)*), the extra carriers (x=1/32) occupy the Fe states at the CBM. They are weakly bound to their nuclei, and behave as shallow donors [25, 26]. Their energy shift from the CBM is so small [17] that we only observe very slight DOS changes with respect to the pristine phase. The situation almost corresponds to a rigid shift of the chemical potential in the frozen pristine DOS and is therefore properly mimicked by a rigid band approximation as often used to access TE properties. Allowing for spin-polarization, this picture is strongly modified: a ferromagnetic (FM) half-metallic phase is energetically favoured, inducing in-gap states (see Fig. 2*(c)*). At x=1/48, those states, mainly of Fe $e_g$ character, are mostly isolated (Fig. 2*(c)*). The spin-splitting is 0.46 eV, with magnetic moments $\mu_{Fe} = 0.28$ $\mu_B$ on the Fe atoms surrounding the impurity (schematically shown in Fig. 1*(b)*). The moment induced on the next-nearest neighbors (Ti) is one order of magnitude smaller, and anti-aligned with Fe; on atoms further away (Sn) it is negligible, showing a strong localization of the magnetization density. At larger Sb concentrations of 1/32 and 1/16 (Fig. 2*(d)* and *(e)*, respectively) the in-gap states start to overlap with the CBM and the spin-splitting decreases to 0.27 eV ($\mu_{Fe} = 0.22$ $\mu_B$) then 0.31 eV ($\mu_{Fe} = 0.23$ $\mu_B$). For the whole range of doping, the integrated magnetization density sums to 1 $\mu_B$ per each

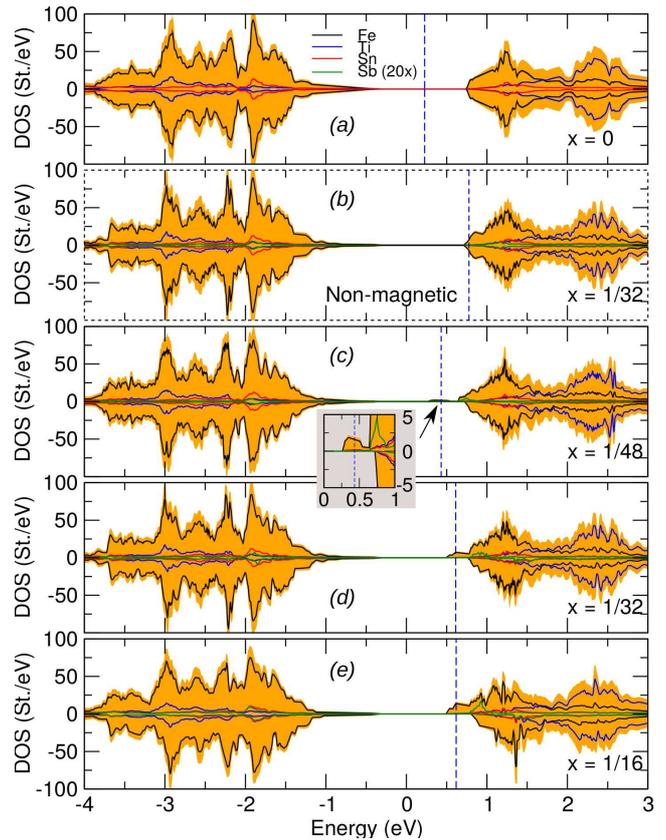

FIG. 2. (Color online) Atom-projected Fe$_2$TiSn$_{Sb}$ DOS (the Sb contributions are magnified 20 times), normalized to the x = 1/16 supercell (B1-WC calculations). $E_F$ is shown as dashed lines. The inset indicates the near-isolated in-gap level for x = 1/48.

Sb atom, corresponding to the integrated DOS of the additional occupied state up to $E_F$. This half-metallic phase is not anticipated when dealing with the rigid band approximation. We obtain similar results in Fe$_2$TiSn$_{As}$: the qualitative change with respect to the rigid band picture is independent of the dopant species [17].

*Chemical effects.* One might wonder if this behavior is also generic to the whole series of Fe$_2YZ$ compounds. From the different band structures shown in Fig. 3 (x=1/32 and x=1/16), we observe that a magnetic instability is present in Fe$_2$TiSn$_{Sb}$, Fe$_2$TiSi$_P$ and Fe$_2$TaGa$_{Ge}$ but absent in Fe$_2$NbGa$_{Ge}$ and Fe$_2$VAl$_{Si}$. As illustrated in Fig. 3 (x=0), the distinct behaviors can be understood based on the electronic band structure of the host matrix, and in particular to the relative position of the Fe and $Y$ $e_g$ bands at the CBM. For Fe$_2$TiSn and Fe$_2$TiSi, the $e_g$ bands of Ti lie well above those of Fe. Under doping, the extra electrons populate the flat band associated to Fe $e_g$ states showing half-metallic spin-splitting. On the contrary, in Fe$_2$NbGa and Fe$_2$VAl, the $e_g$ bands of Nb and V lie well below those of Fe. The extra electrons therefore populate the highly disper-
2

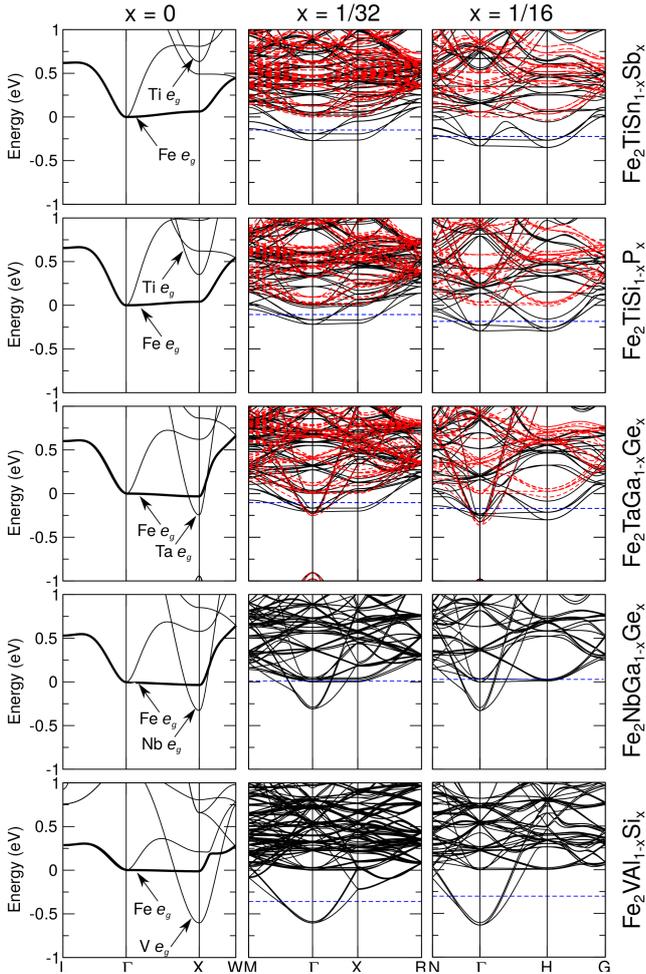

FIG. 3. (Color online) Spin-resolved electronic band structures (B1-WC calculations) of explicitly doped $Fe_2YZ_A$ compounds for distinct doping concentrations x in the associated irreducible Brillouin zone ($Fm\bar{3}m$, $Im\bar{3}m$ and $Pm\bar{3}m$, respectively, for x = 0, 1/32 and 1/16). The zero of energy is set to the bottom of the Fe $e_g$ band at $\Gamma$. Blue dashed line: Fermi energy $E_F$. Red dashed lines: minority spin channel.

sive $Y$ $e_g$ band, and no magnetic transition is observed. $Fe_2TaGa$ is in an intermediate situation, with Fe and Ta $e_g$ states closer in energy, so that at the investigated doping concentrations both are occupied. The system exhibits a magnetic instability, but the energy difference between FM and NM phases is smaller than for $Fe_2TiSn$ and $Fe_2TiSi$. This contradiction between using flat bands to increase the PF and the risk of magnetic instabilities adds yet another constraint to the optimization of TE materials, which has not been appreciated so far in the literature.

*Origin of the magnetic instability.* From the above, it appears that a magnetic instability takes place when doping electrons start populating the localized Fe $e_g$ states. In order to validate this explanation and explore further the origin of the magnetic instability, we perform a simple numerical experiment using the alternative GGA+U approach. Considering $Fe_2TiSn$ as a reference compound, we artificially tune the amplitude of the $U_{Ti}$ parameter (from 0.0 to 5.6 eV, see Ref. [17]) in order to modify the relative position of Fe and $Y$ (Ti) $e_g$ levels and mimic the distinct band structures of the whole series of $Fe_2YZ$ compounds reported in Fig. 3 without explicitly changing the cations. The different $e_g$ band arrangements illustrated in Fig. 4(a-d) (top row) properly reproduce the different regimes identified in Fig. 3, and are then used as hosts for implicit doping achieved by adding extra electrons and a compensating positive background. The spin-projected DOS at $E_F$ and the total cell magnetization are reported in Fig. 4(a-d) (bottom row) as a function of the carrier concentration. In order to determine the doping windows where itinerant electron magnetism (typical of intermetallic alloys [27, 28]) is expected to dominate [29–31], the Stoner criterion [31, 32] is also evaluated: a FM state is favored when the product between the NM DOS at $E_F$ and the energy needed to flip a spin ($\Delta E_{ex}$) is larger than one ($St = DOS(E_F) \cdot \Delta E_{ex} > 1$).

When the $Y$ $e_g$ states lie significantly below the Fe $e_g$ states (panels a-b), the system is NM at small carrier concentrations (*i.e.* when doping electrons occupy exclusively $Y$ $e_g$ states) and then becomes FM when $E_F$ touches the Fe $e_g$ states. This also coincides with $St > 1$ so that the appearance of magnetism is compatible with a Stoner instability. When the Fe $e_g$ states lie below the $Y$ $e_g$ states (panels d), doping electrons immediately occupy Fe $e_g$ states and the system is always magnetic, independently of the Stoner criterion. In the intermediate case where the $Y$ $e_g$ state minimum is below but close to the Fe $e_g$ states (panel c), the system is initially NM and becomes FM as soon as Fe $e_g$ states start to be populated. This shows that, although different regimes might exist depending on the value of $St$, the appearance of magnetism is not always the result of a Stoner instability, but rather intrinsic to the Fe 3d $e_g$ states, which are strongly localized and experience robust magnetic exchange interactions. This means that in compounds like $Fe_2TiSi$ or $Fe_2TiSn$, a magnetic ground state cannot be avoided, even at small carrier concentrations where $St < 1$. It also suggests that substituting the strongly localized 3d orbitals of Fe by the more delocalized 4d or 5d orbitals of Ru or Os might delay the appearance of magnetism.

To test this, we consider $Ru_2ZrSn$ and $Os_2HfSn$ Heusler compounds, which have not been synthesized to our knowledge. As illustrated in Fig. 4(e-f), in these cases magnetism is no longer tied to the occupation of the d-states: it results from a proper Stoner instability and appears only when $St \approx 1$ [33], leaving a wide range of carrier concentrations for which Zr or Hf d-states are partially occupied but the system remains NM. For heavy cations one could expect that the SOC (neglected for Fe above) might play an important role, and we in-clude it in the Ru and Os calculations. As can be seen in Fig. 4(e), it has no significant effect in the case of $Ru_2ZrSn$. For $Os_2HfSn$ however, it changes the band structure more substantially, and suppresses the magnetic instability in the whole range of carrier concentrations explored in Fig. 4(f).

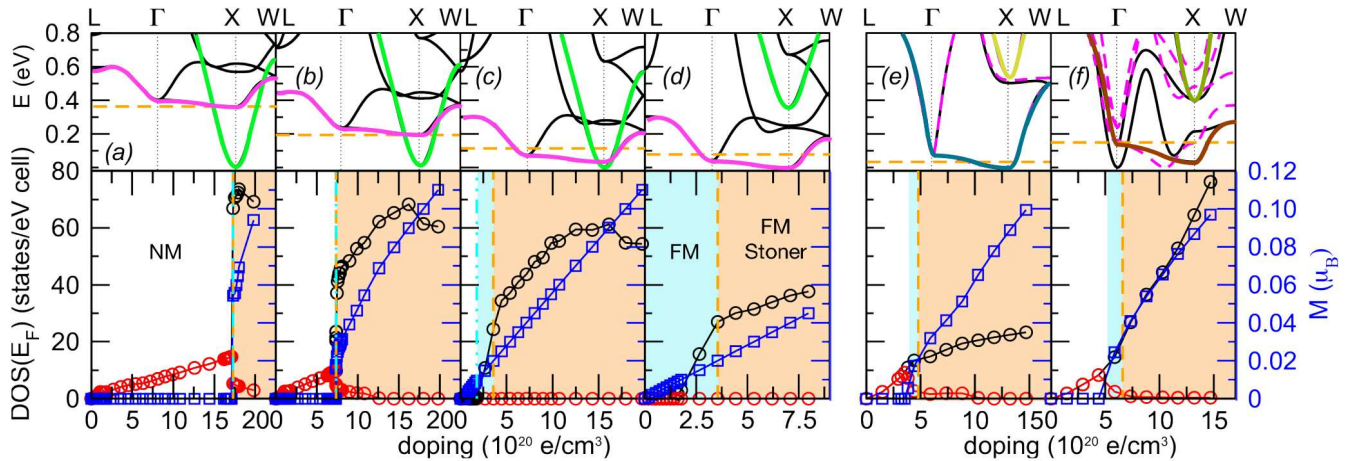

FIG. 4. (Color online) Top row - Electronic band structures (GGA+$U$ calculations) of the pristine phase of *implicit* $Fe_2TiSn$ at distinct $U_{Ti}$ mimicking the different $Fe_2YZ$ compounds of Fig. 3 (see text) : *(a)* $U_{Ti}$ =0.0 eV, *(b)* $U_{Ti}$ =1.4 eV, *(c)* $U_{Ti}$ =2.6 eV and *(d)* $U_{Ti}$ =5.6 eV. Electronic band structures (GGA+$U$ calculations) of the pristine phase *(e)* $Ru_2ZrSn$ *(f)* $Os_2HfSn$ (dashed-line magenta bands include SOC). Bottom row - Related evolution of the projected DOS at $E_F$ for up (black circles) and down (red circles) spins and total magnetization (blue squares) in terms of implicit electron doping. The critical doping needed to start populating the Fe $e_g$ levels is identified by a cyan double-dot dashed line. The critical doping for which $St > 1$ is identified by an orange dashed line (also in the top row). The white, blue and orange areas identify the non-magnetic (NM), regular ferromagnetic (FM) and ferromagnetic Stoner (FM-Stoner) regimes.

*Thermoelectric properties.* Having demonstrated the appearance of a magnetic instability under doping, it is now important to clarify its consequences on transport and TE properties. To this end, we compare the evolution of the PF as a function of the chemical potential, $\mu$. Our calculations rely on Boltzmann transport theory and the rigid band approximation [17] using either the electronic band structure of the pristine phase or that of the doped system in the NM and eventually FM configurations. For the purpose of comparison, in the latter cases, the zero of $\mu$ was defined in order to align deep energy levels on those of the pristine phase. Ideally, calculations at each $\mu$ should rely on the band structure at the related carrier concentration. Still, comparing here full curves obtained from the rigid band structure at different carrier concentrations allows us to probe the quality of the rigid band approximation.

The results for two representative cases, $Fe_2TiSn_{Sb}$ and $Fe_2NbGa_{Ge}$ (x = 1/16 at 300 K), are shown in Fig. 5*(a)* and *(b)* respectively. A vertical line locates the position of $E_F$ when considering the band structure of a doped system. For $Fe_2NbGa_{Ge}$, which remains NM at x = 1/16, the shape of the PF remains almost unchanged when using the band structure of the pristine or explicitly doped phase, with just a slight reduction of the main peak by a factor 1.3. This confirms that, as already shown in Fig. 2, doping does not significantly affect the band structure so that the rigid band approximation provides a realistic estimate of PF in that case. This remains true for $Fe_2TiSn_{Sb}$ when considering the NM phase. However, when considering the band structure of the FM ground state at x = 1/16, the PF changes drastically and the main peak shifts and drops by a factor of 4.3. This highlights that spin-splitting is strongly detrimental to the PF. This can be related to the sensitivity of $S$ to modifications of the band structure and chemical potential: although the number of additional carriers is fixed (one electron per site), fewer states in a range of $k_BT$ around $E_F$ contribute to transport, causing the decrease of $S$ [17]. Such an effect cannot be anticipated when considering the pristine (NM) phase and the rigid band approximation.

In Fig. 5*(c)*, we report the temperature dependence of the PF for various doped $Fe_2YZ_A$ systems at a dopant concentration of x = 1/16, using the band structure under explicit doping and for the magnetic ground state. Although the values are reduced compared to those previously reported [9], relatively large PF can still be observed. The largest values are for $Fe_2NbGa_{Ge}$ (which remains NM) and $Fe_2TaGa_{Ge}$ (which is at the limit of FM). But, even the PF of $Fe_2TiSi_P$, although significantly reduced by the FM instability, remains sizable and larger than that of $Fe_2VAl_{Si}$, confirming the interest of $Fe_2YZ$ compounds for TE applications [34]. Moreover, this makes the worst hypothesis that compounds with a FM ground state remain FM at operating temperatures, which might not necessarily be the case. We generally expect the exchange splitting to decrease with $T$, which together with the enhanced spin fluctuations and carrier-magnetic interactions at high $T$, could further improve the TE properties of the doped Heusler with magnetic instabilities [34, 35].

As previously discussed, substituting Fe by Ru or Os is a way to delay, or even suppress, the emergence of the detrimental magnetic instability, enlarging the doping region in which the system remains NM. In Fig. 5(d), we report the PF of hypothetical $Ru_2ZrSn$ and $Os_2HfSn$ at 300 K. For $Ru_2ZrSn$, relying on the band structure of the pristine phase we predict a large PF of $16.1 \times 10^{-3}$ W/m K$^2$. This result is confirmed from calculations with the band structure at a carrier concentration of $2.5 \times 10^{20}$ cm$^3$, which remains in the NM regime. At larger carrier concentrations around n = $10.0 \times 10^{20}$ cm$^3$, the PF is significantly reduced when reaching the FM regime. For $Os_2HfSn$, SOC can no longer be neglected and suppresses the magnetic instability in the whole range of studied carrier concentrations. In that case, although the band structure is significantly modified by SOC, the PF can still reach extremely large values of $22.3 \times 10^{-3}$ W/m K$^2$ (up to 45.5 when neglecting SOC - not shown). [36]. Although Ru and Os are expensive and likely not a scalable solution for TE applications, this confirms that larger PF can be achieve using $4d$ and $5d$ elements.

*Conclusions.* From calculations on $Fe_2YZ$ full Heusler compounds, under explicit doping conditions compatible with thermoelectric applications, we have shed light on a previously overlooked magnetic instability, detrimental to their TE properties. At a time where the discovery of new TE materials relies more and more on high-throughput searches based on the rigid band approximation [7, 37], our study shows that we must remain extremely careful: although relying on the band structure of the pristine phase will often provide a good estimate, further validation under explicit doping should be systematically performed. The magnetic instability of $Fe_2YZ$ compounds is assigned to the strong localization of the Fe $3d$ states and can be delayed or even suppressed using $4d$ and $5d$ elements. Moreover, even when the system becomes magnetic, the loss of carriers contributing to transport is not always dramatic, and can maintain a large PF compared to other prototypical TE systems (PF $\sim$ 3 - 4 mW/m K$^2$ at 300 K in $Fe_2VAl$ [38, 39] or PF $\sim$ 4 - 5 mW/m K$^2$ in PbTe [40]). More generally, the electronic band structure engineering highlighted in this work (manipulation of in-gap states, ferromagnetism and/or half-metallicity) also opens exciting perspectives for spintronic and spin-caloritronic applications [41, 42]. The exploitation of charge, spin and heat transport with fully spin-polarized carriers, for example in the spin-Seebeck or spin-Nernst effects, together with cheap and abundant atomic components in the full-Heusler alloys,

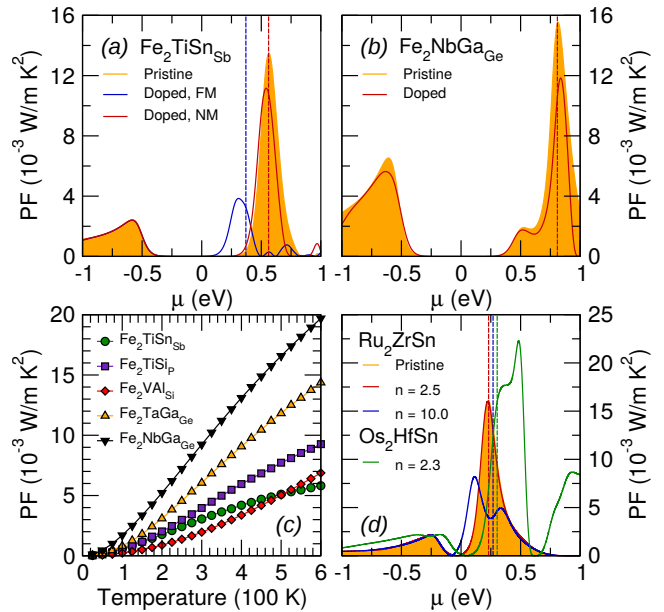

FIG. 5. (Color online) Top panels - Spectral thermoelectric PF of *(a)* $Fe_2TiSn_{Sb}$ and *(b)* $Fe_2NbGa_{Ge}$ with respect to $\mu$, at 300 K, in the rigid band approximation, and for explicit doping with x = 1/16 (B1-WC calculations). Bottom panels - *(c)* Evolution of the PF with respect to the temperature (B1-WC calculations) for various explicitly doped $Fe_2YZ_A$ compounds at x = 1/16 (the constant relaxation time approximation overestimates the PF values at high T, near 600 K [9]). *(d)* Evolution of the PF in terms of $\mu$ (GGA+$U$ calculations) at 300K for implicitly doped $Ru_2ZrSn$ and $Os_2HfSn$ ($n$ values are expressed in $10^{20}$ cm$^{-3}$). In panels (a), (b) and (d), the vertical dashed lines locate the position of $E_F$ of the doped phases.

might be a starting point for low-cost thermo-magnetic applications.

We thank D. Singh, E. Bousquet and A. Mercy for fruitful discussions. F.R., M.J.V. and Ph.G. were supported by the European Funds for Regional Developments (FEDER) and the Walloon Region in the framework of the operational program "Wallonie-2020.EU" (project: Multifunctional thin films/LoCoTED). S.L., M.J.V., and Ph.G. were supported by the ARC project AIMED 15/19-09 785. Calculations have been performed on the Belgian CECI facilities funded by F.R.S-FNRS Belgium (Grant No 2.5020.1) and Tier-1 supercomputer of the Fédération Wallonie-Bruxelles funded by the Walloon Region (Grant No 1117545). M.J.V. acknowledges a FNRS sabbatical grant, hosted by the Catalan Institute of Nanoscience and Nanotechnology Barcelona. F.R. and S.L. contributed equally to this work.

---

* F.R. and S.L. contributed equally to this work.

# SUPPLEMENTARY MATERIAL:
# Magnetic instabilities in doped $Fe_2YZ$ full-Heusler thermoelectric compounds

Sébastien Lemal,[1] Fabio Ricci,[1] Daniel I. Bilc,[2] Matthieu J. Verstraete,[1,3] and Philippe Ghosez[1]

[1]*Physique Théorique des Matériaux, Q-MAT, CESAM,*
*Université de Liège (B5), B-4000 Liège, Belgium*
[2]*Faculty of Physics, Babeş-Bolyai University, 1 Kogălniceanu, 400084 Cluj-Napoca, Romania*
[3]*European Theoretical Spectroscopy Facility, http://www.etsf.eu/*
(Dated: November 18, 2019)

## Motivations

The main purpose of the $n$-type doped $Fe_2YZ_{1-x}A_x$ ($Fe_2YZ_A$) compounds is to have the $A$ impurities acting as shallow donors so that they do not perturb the states close to the band gap keeping thus other properties of the pristine compounds unaffected. The shift of the chemical potential on the Fe $e_g$ band is done by substituting part of the $Z$ atoms with $A$ atoms so that the atomic number goes from $Z_A = Z_Z + 1$. Hence, the substitutional atom is similar to the $Z$ atom in term of size, mass and electronic structure just with one additional electron. In these conditions, we observe wether the Fermi level ($E_F$) is shifted at the maximum of the spectral power factor [1].

## Technical details

DFT calculations were performed with the CRYSTAL code [2, 3]. The B1 Wu-Cohen [4] (B1-WC) hybrid functional has been used for all calculations. Doping was modelled by the means of cubic and tetragonal supercells, giving a range of doping from $3.8 \times 10^{20}$ cm$^{-3}$ to $1.5 \times 10^{21}$ cm$^{-3}$, falling into the range of doping needed to maximize the power factor as identified in Ref. 1. The following compounds are studied: $Fe_2TiSn_{Sb}$, $Fe_2TiSi_P$, $Fe_2VAl_{Si}$, $Fe_2TaGa_{Ge}$ and $Fe_2NbGa_{Ge}$. In all cases, the compositions x = 0, 1/32 and 1/16 have been considered. For the specific case of $Fe_2TiSn_{Sb}$, we also investigated the doping value x = 1/48 with a $2 \times 2 \times 3$ supercell.

The basis set used are taken from Ref. 5 for Fe, Ref. 6 for Ti, Ref. 7 for Sn, Sb, Nb, Al and Ta, Ref. 8 for V and Ge, Ref. 9 for Ga, and Ref. 10 for Si. Spin-polarization is considered: an initial magnetic moment of 1 $\mu_B$/u.c. is imposed to the unit cell during the first 3 steps of the self-consistent cycle. Different Monkhorst-Pack [11] meshes of k-points were used: i) a $9 \times 9 \times 9$ mesh was used for the structural relaxation of the undoped unit cells; ii) a $5 \times 5 \times 5$ mesh was used for the structural relaxation of the doped supercells; iii) a $10 \times 10 \times 10$ mesh was used for the computation of the electronic properties; and iv) a $32 \times 32 \times 32$ mesh was used for the computation of the thermoelectric properties after interpolation. The energy convergence criterion was fixed to $10^{-9}$ Ha. For relaxation, we fixed a threshold of $3 \times 10^{-4}$ Ha/Bohr on the root-mean square values of energy gradients and of $1.2 \times 10^{-3}$ Bohr on the root-mean square values of atomic displacements. A temperature smearing of the Fermi surface was set to $3.2 \times 10^{-4}$ Ha. For transport properties, the BoltzTraP [12] code was used, which performs calculation within the constant relaxation time approximation (CRTA). The constant relaxation time $\tau = 3.4 \times 10^{-14}$ s is taken from Ref. 1. Further information on its estimation is given in the Supplementary Materials of Ref. 1.

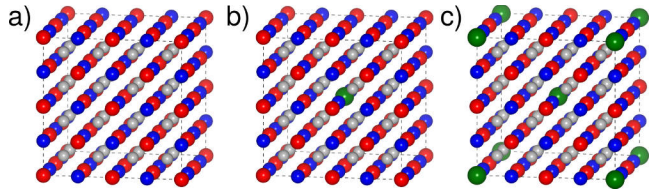

FIG. 1. Structure of $Fe_2YZ$ compounds with, from left to right, no impurity, one impurity and two impurities (green) in the $2 \times 2 \times 2$ supercell.

The DFT+$U$ simulations using the ABINIT code [13] were performed within the PBE [14] flavour of the generalized gradient approximation (GGA). We make use of the $U$ correction in order to treat the electronic correlations on the transition metal atoms $d$ orbitals [15] and we self-consistently determined the Hubbard-like $U$ correction using the linear response formalism [16], with $5.0 \times 10^{-2}$ eV degree of convergence on the $U$ values, corresponding to a the lattice parameter optimization tolerance of the order of $10^{-3}$ Å. We used projected augmented-wave (PAW) pseudopotentials [17] taken from the JTH table [18] and in order to achieve a satisfactory degree of convergence ($\sim$0.01 meV energy differences) the plane wave expansion has been truncated at a cutoff energy of 653 eV and the integrations over the Brillouin Zone were performed considering $20 \times 20 \times 20$ uniform Monkhorst and Pack grid [11].

In this framework, we obtain (i) lattice parameters in perfect agreement with available experiments ($\Delta a < 5$ ‰, see Tab. I); (ii) similar static and dynamic charges on each atomic site with respect to B1-WC calculations (not shown) and (iii) band structures reasonably similar to B1-WC (similar band gaps, and states at the valence and conduction edges, see Fig. 2). It is worth noticing however that, just for $Fe_2TiSn$ and similarly for $Fe_2TiSi$ (not shown), using the self-consistent $U_{Ti}$ results in a Ti $e_g$ (dispersive) band, too low in energy, with a minimum touching the Fe $e_g$ (flat) band. As illustrated in Fig. 3, using a larger $U_{Ti} = 5.60$ eV permits to recover a better agreement in the bottom of the conduction band, yielding thermoelectric properties similar to those determined from the B1-WC band structure.

| $Fe_2YZ$ | $U_{Fe}$ (eV) | $U_Y$ (eV) | $E_g$ (eV) | $a$ (Å) | $a_{EXP}$ (Å) | Ref. |
|---|---|---|---|---|---|---|
| $Fe_2TiSn$ | 5.09 | 2.62 | 1.29 | 6.069 | 6.074 | [19] |
| $Fe_2TiSi$ | 5.02 | 2.47 | 1.41 | 5.714 | 5.720 | [20] |
| $Fe_2VAl$ | 5.02 | 4.86 | 1.09 | 5.733 | 5.761 | [21] |
| $Fe_2TaGa$ | 5.02 | 1.28 | 1.35 | 5.929 | - | - |
| $Fe_2NbGa$ | 5.03 | 1.55 | 1.02 | 5.934 | - | - |
| $X_2YZ$ | $U_X$ (eV) | $U_Y$ (eV) | $E_g$ (eV) | $a$ (Å) | $a_{EXP}$ (Å) | Ref. |
| $Ru_2ZrSn$ | 2.90 | 1.09 | 0.17 | 6.479 | - | - |
| + SOC | 2.88 | 1.03 | 0.15 | 6.479 | - | - |
| $Os_2HfSn$ | 2.67 | 0.99 | 0.59 | 6.484 | - | - |
| + SOC | 2.82 | 0.25 | 0.14 | 6.483 | - | - |

TABLE I. GGA+$U$ study: self-consistently determined $U$ for the transition metal atoms (Fe and $Y$ sites) in the overall compounds; obtained energy gap ($E_g$) and related optimized and experimentally available lattice parameters $a$.

In addition, the effect of the spin-orbit coupling (SOC) has been checked on the $4d$ and $5d$ transition metals computing again the self-consistent $U$ values, the equilibrium lattice parameters and the electronic properties. The results are reported on Tab. I and the band structure differences are shown in Fig. 4, 5, 6, 7, 8, 9 and 10.

## Non-magnetic shallow donor levels

As we have seen in the main text (Fig. 1(b)), the mechanism driving the doping consequences in the non-magnetic constrained phase is shallow donor-like. A further proof for this behaviour can be given exploiting the Effective Mass Theory [22–24]. In this approach, the binding energy of the donor level $E_D^b$ with respect to the conduction band minimum (CBM) and the spatial extension $a_D$ of the related wave function can be expressed as:

$$E_D^b = \frac{m^*}{m} \frac{hc\, R_\infty}{\varepsilon_\infty^2} \quad (1)$$

$$a_D = \frac{m}{m^*} \varepsilon_\infty\, a_B \quad (2)$$

where $m$ ($m^*$) is the mass (effective mass) of the additional electron, $\varepsilon_\infty = 25.55$ is the crystal dielectric constant (estimated within GGA+$U$), $R_\infty$ is the Rydberg constant with $hc\, R_\infty = 13.61$ eV and $a_B = 0.53$ Å, the Bohr radius. In the case of $Fe_2TiSn$, the effective masses related to the dispersive ($m_l$) and flat ($m_h$) bands, were estimated by Bilc and coworkers [1] to be $m_l = 0.3\, m$ and $m_h = 26\, m$. The two separated contributions give:

$$E_D^b(m_l) \approx 4.6 \text{ meV and } a_D(m_l) \approx 45.0 \text{ Å}\, ; \quad (3)$$

$$E_D^b(m_h) \approx 540.0 \text{ meV and } a_D(m_h) \approx 0.5 \text{ Å}\, . \quad (4)$$

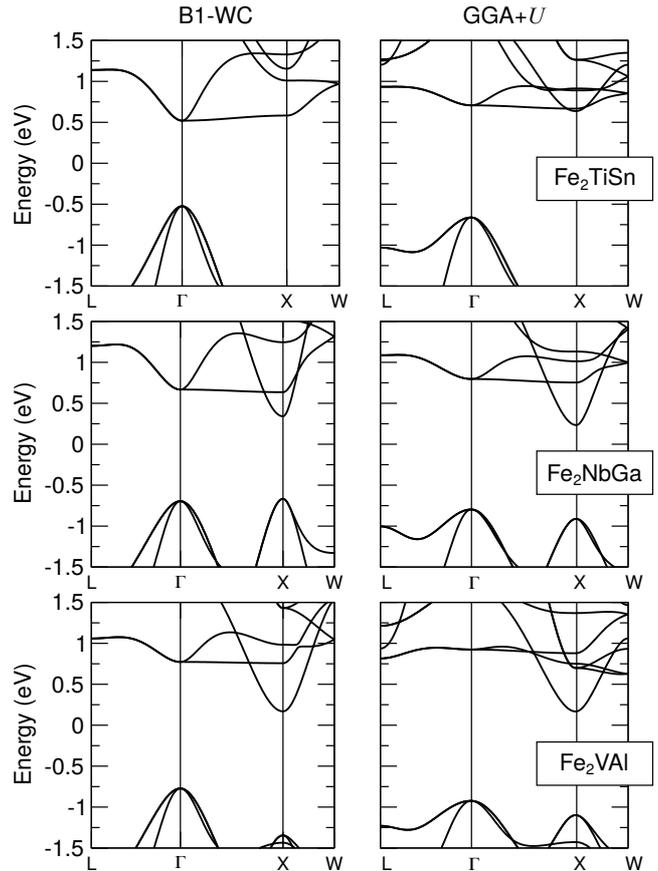

FIG. 2. Band structures of $Fe_2TiSn$, $Fe_2NbGa$ and $Fe_2VAl$ as calculated from B1-WC and GGA+$U$ with self-consistent $U$'s.

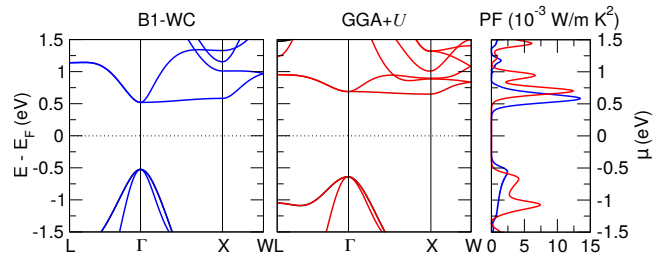

FIG. 3. Band structures of $Fe_2TiSn$ as calculated from B1-WC and GGA+$U$ ($U_{Fe} = 5.09$ eV, $U_{Ti} = 5.60$ eV). The spectral power factors as calculated from both band structures, in the rigid band approximation, are shown on the right. Both methods reproduces the peak arising from the Fe $e_g$ band at the CBM.

It is clear at this point that the results in Eq. (3) are in agreement with the picture given for Fig. 1($b$) where the shallow donor mechanism induces a doping level practically incorporated to the CBM and causing a rigid-band-like shift of the chemical potential. On the contrary, in case of flat bands the doping level would be completely isolated and far from the CBM, as quantified by Eq. (4). However, as explained in the main text, the electronic localization is driven by the exchange interaction and not by the (shallow donor) nature of doping.

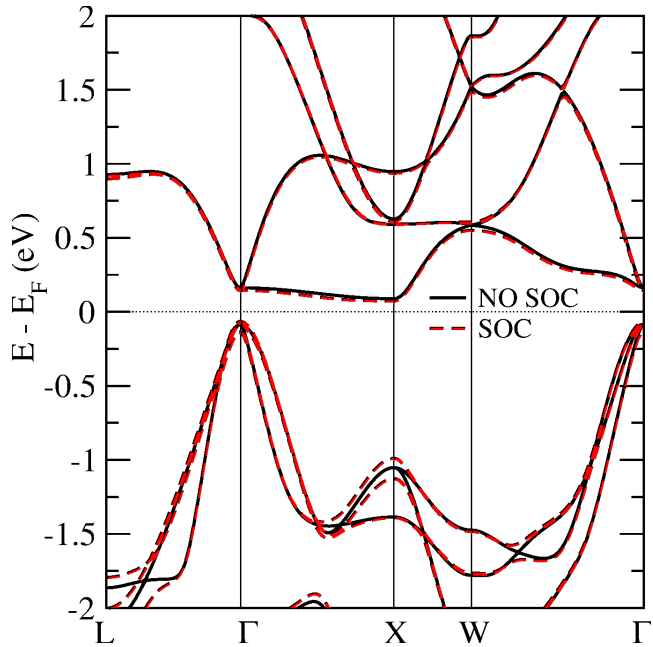

FIG. 4. Ru$_2$ZrSn band structures accounting for or not the SOC interaction, as calculated from GGA+$U$.

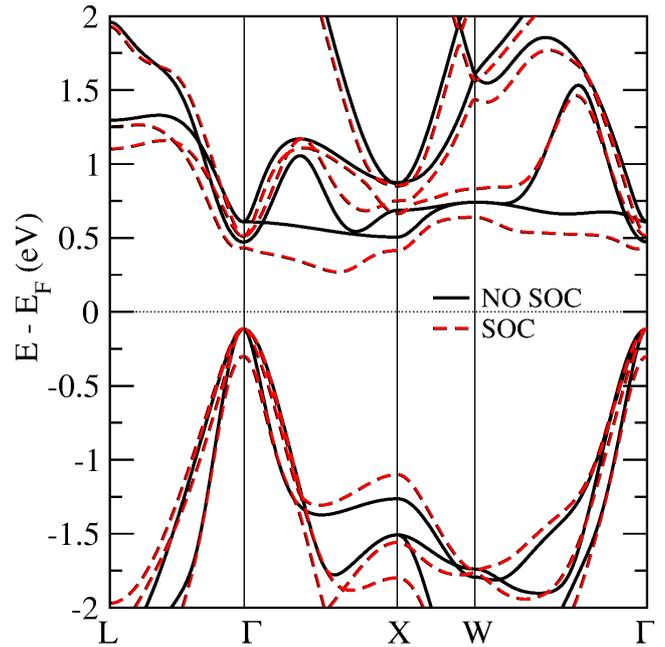

FIG. 6. Os$_2$HfSn band structures accounting for or not the SOC interaction, as calculated from GGA+$U$.

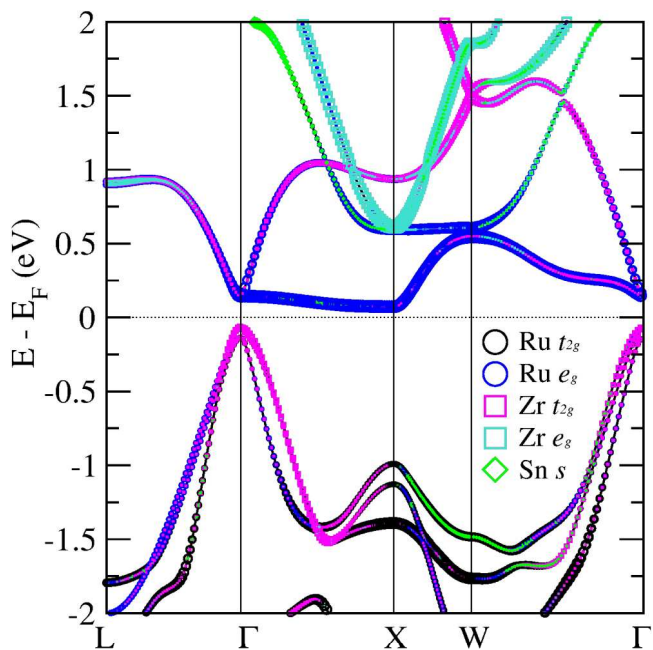

FIG. 5. Ru$_2$ZrSn orbital-weighted band structures including SOC interaction, as calculated from GGA+$U$.

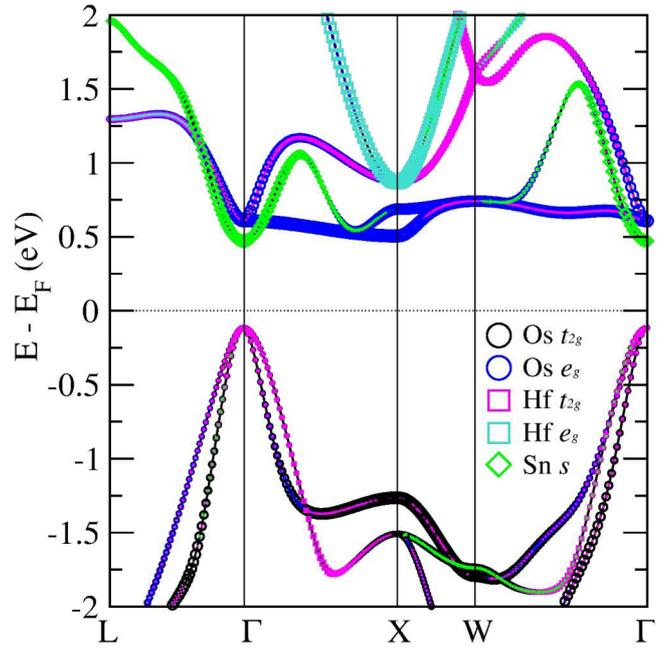

FIG. 7. Os$_2$HfSn orbital-weighted band structures without SOC interaction, as calculated from GGA+$U$.

### Magnetic phases in doped compounds

The magnetization energies $\Delta E$ of the doped Fe$_2YZ_A$ phases (difference between NM and FM total energies)

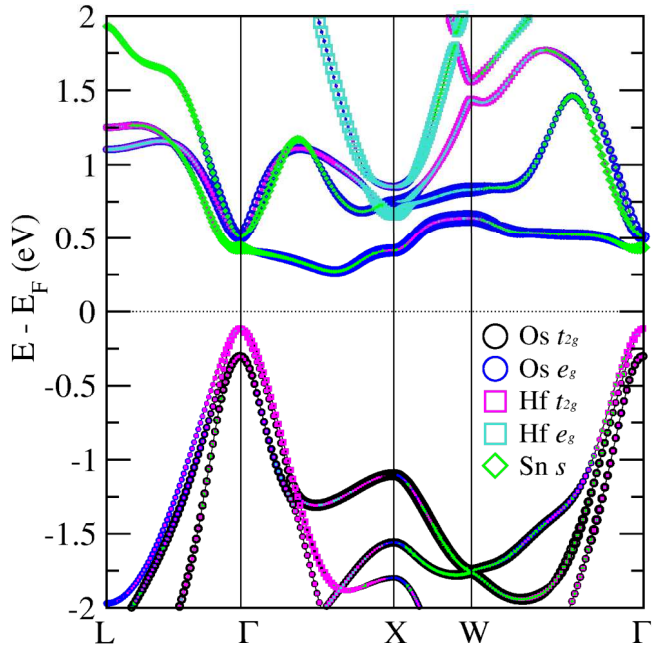

FIG. 8. $Os_2HfSn$ orbital-weighted band structures including SOC interaction, as calculated from GGA+$U$.

| $Fe_2YZ_A$ | x | $\Delta E$ (meV) |
|---|---|---|
| $Fe_2TiSn_{Sb}$ | 1/32 | 75 |
|  | 1/16 | 166 |
| $Fe_2TiSi_P$ | 1/32 | 51 |
|  | 1/16 | 105 |
| $Fe_2VAl_{Si}$ | 1/32 | 0 |
|  | 1/16 | 0 |
| $Fe_2TaGa_{Ge}$ | 1/32 | 37 |
|  | 1/16 | 98 |
| $Fe_2NbGa_{Ge}$ | 1/32 | -34 |
|  | 1/16 | -7 |

TABLE II. Magnetization energies ($\Delta E = E_{NM} - E_{FM}$) of the 128-atoms supercells of the doped $Fe_2YZ_A$ at different compositions, as calculated from B1-WC: a positive value indicates that the FM phase is more stable than the NM phase.

are given in Table II. From here, it is evident the stability of the FM phase for $Fe_2TiSn_{Sb}$, $Fe_2TiSi_P$ and $Fe_2TaGa_{Ge}$. Concerning the $Fe_2VAl_{Si}$, no magnetic phase can be stabilized while for $Fe_2NbGa_{Ge}$ the NM phase is the groundstate.

In addition to the FM phase, different antiferromagnetic (AFM) configurations have been explored with both B1-WC and GGA+$U$ methods. However, for each one of these the charge density cannot be converged, hence no related results can be shown at present. This behaviour, nevertheless, suggests that the FM phase is robust, as expected from the large exchange-splittings found in the band structures. We also stress that the density of dopants (fixed by the choice of the supercell sizes) guaranties that the carriers density satisfies the Stoner criterion for itinerant-electron magnetism.

### Charge localization in doped systems

The strong donor charge density surrounding the defect that typically accompanies the ferromagnetic (FM) phase transition can be seen from the charge density projected on the occupied conduction states (Fig. 11). This behaviour is the consequence of the highly localized nature of the Fe $e_g$ orbitals accommodating the additional electrons. In the case of $Fe_2TiSn_{Sb}$ (similar to $Fe_2TiSi_P$ due to the similar band structure, see main text), the shape of the projected electron density corresponds exactly to the magnetization density, with the whole form of $e_g$ orbital near the Sb impurity. In the case of $Fe_2NbGa_{Ge}$, the additional charge is fully delocalized over the Fe and Nb atoms with a smaller filling. In the case of $Fe_2VAl_{Si}$, the charge delocalizes all over the V atoms as expected from the band structure (see Fig. 3 in the main text). $Fe_2TaGa_{Ge}$ is in an intermediate case between $Fe_2TiSn_{Sb}$ and $Fe_2NbGa_{Ge}$ (not shown). Interestingly, the localization effect and the magnetic phase transitions also appear upon injection of additional electrons in the pristine structures with a compensating positively charged background: this suggests a strictly electronic origin of these phenomena.

### Nature of the donors

We investigated the role of the dopants species by computing the band structure of $Fe_2TiSn$ doped with As instead of Sb, shown in Fig. 12. An exchange splitting of 0.25 eV is witnessed between the minority and majority spin population near the conduction band minimum, similar to what is observed for $Fe_2TiSn_{Sb}$. Hence, we do not expect the nature of the donor to play a significant role in the observed magnetic properties. This is further justified in the following Sections.

### Artificial doping within B1-WC

To disentangle the atomic size effect (due to the different size of the dopant with respect to the substituted atom in the pristine phase) from the electron doping itself, the localization of additional carriers is also witnessed in a $2 \times 2 \times 2$ supercell of $Fe_2YZ$, by adding a single electron and a compensating charged background at fixed cubic geometry (in a Jellium-like picture), in order to mimic the effect of the $x = 1/32$ substitution. The corresponding band structures for (a) $Fe_2TiSn$, (b)

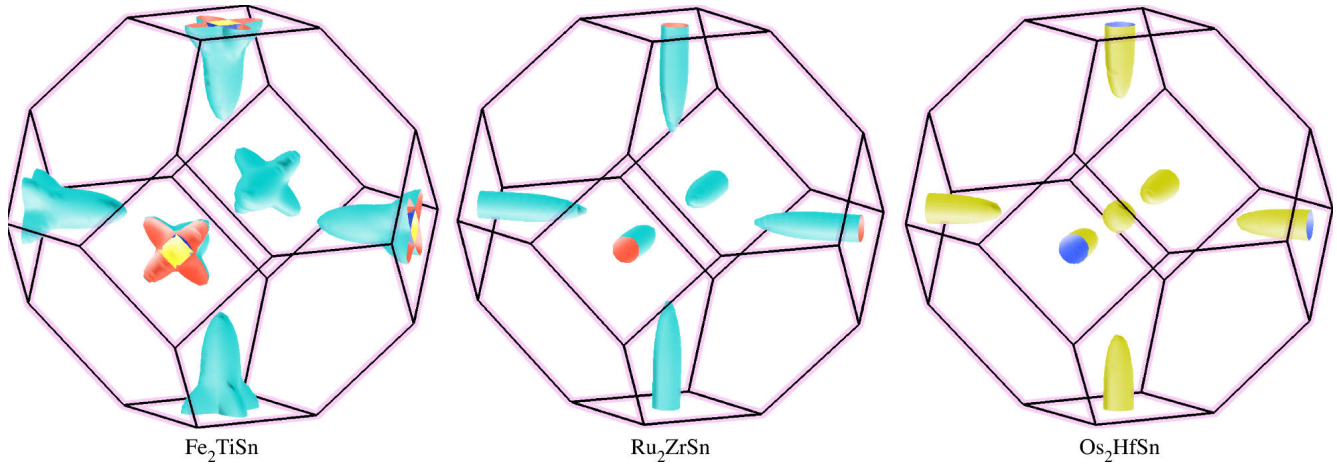

FIG. 9. Fermi Surfaces without SOC interaction, as calculated from GGA+$U$.

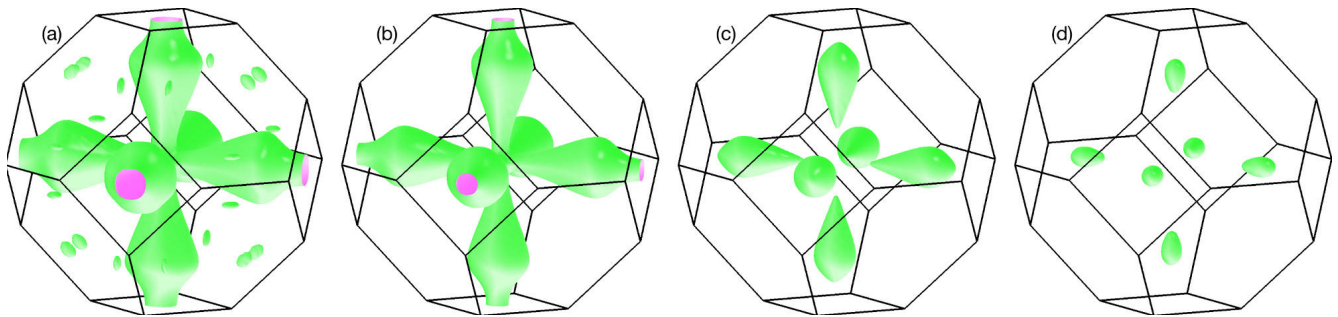

FIG. 10. Os$_2$HfSn Fermi Surfaces including SOC interaction at: (a) major PF peak; (b) 90% of major PF peak; (c) PF flat region; (d) 90% of PF flat region, as calculated from GGA+$U$.

Fe$_2$NbGa and (c) Fe$_2$VAl are shown in Fig. 13. In the case of Fe$_2$TiSn, $E_F$ is shifted toward the conduction band, and the Fe $e_g$ states evidence an exchange splitting (200 meV/cell), similarly to the explicitly doped case. On the contrary, no exchange splitting occurs for Fe$_2$NbGa and Fe$_2$VAl again in agreement with the explicitly doped procedure. The role of the $Y$ $e_g$ orbitals is fundamental in this phenomenon: the population belonging to the flat region (Fe $e_g$, with high effective mass and low mobility) is expected to localize in real space and its strong electronic exchange interaction favours the spin-splitting, whereas the one on the highly dispersive band ($Y$ $e_g$, with low effective mass and high mobility), delocalizes and disadvantages the magnetic phase.

The interplay between magnetization and localization can be studied through the specific case of Fe$_2$TiSn as illustrated in Fig. 14, where the band structures are plotted for the $2 \times 2 \times 2$ supercell doped with (a) one electron in its FM ground-state (same Figure as Fig. 13(a)), (b) for the non-magnetic (NM) constrained phase and (c) the pristine phase. The band profile in Fig. 14(b) and Fig. 14(c) cases mostly only differs by the position of the $E_F$. This shows that the impurity atom Sb is not even needed for the exchange-splitting to manifest (with a gain of energy of 44 meV with respect to the NM phase), suggesting an electronic origin.

The associated electron densities for the added carriers are shown in Fig. 15(a) for the FM ground state and Fig. 15(b) for the NM constrained phase. In the first case, localization occurs: the pattern of electron density is the same as in the explicitly doped case (Fig. 15(c), while in the second case, the charge is completely delocalized over the Fe, as expected from the rigid band approximation. As a consequence of these results, the localization of carriers strictly shows an electronic origin and is associated with a magnetic instability.

We verified that the electronic nature of the magnetic instability appearing in Fe$_2$TiSn, Fe$_2$TiSi and Fe$_2$TaGa is Stoner-type [25, 26]. For this purpose, the Stoner criterion is evaluated for all the doped compounds, accounting for the exchange splitting $\Delta E_{ex}$ and the non-magnetic phase DOS at $E_F$, DOS$^{(NM)}(E_F)$. According to the Stoner model of the ferromagnetism, a magnetic instability occurs when:

$$St = \text{DOS}^{(\text{NM})}(E_F) \cdot \Delta E_{\text{ex}} > 1 \qquad (5)$$

The results obtained are listed in Tab. III and, very interestingly, each compound showing a FM phase fulfils the criterion. For Fe$_2$VAl and Fe$_2$NbGa, $St < 1$, as expected from the DFT results. These results suggest that the magnetic phase transition as a function of the doping is due to a Stoner instability in $Fe_2TiSn$, $Fe_2TiSi$ and $Fe_2TaGa$.

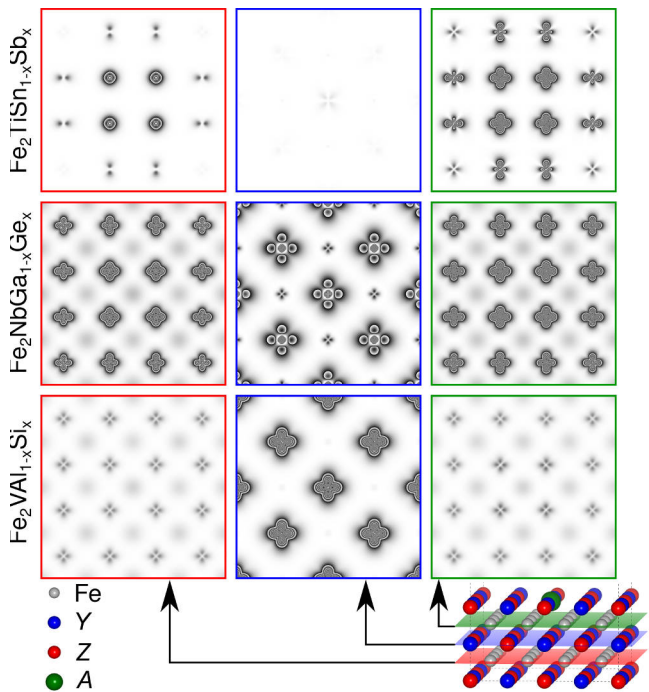

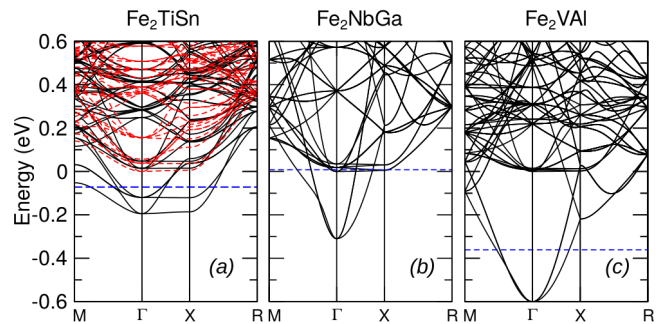

FIG. 13. Ground state band structure of $Fe_2YZ$ computed on $2\times 2\times 2$ supercells (at fixed geometry) injecting one additional electron, as calculated using B1-WC. Dashed line: $E_F$; black- (red-) line: spin-up (spin-down) channel.

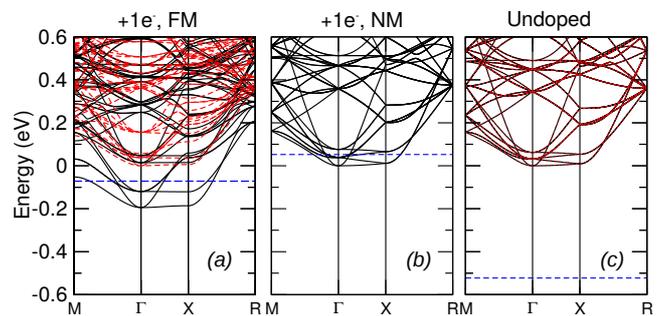

FIG. 14. Band structure of $Fe_2TiSn$ computed on $2 \times 2 \times 2$ supercell (at fixed geometry), as calculated using B1-WC: *(a)* doped with one additional electron in its FM ground-state; *(b)* doped with one additional electron in the NM constrained phase; *(c)* undoped. Dashed line: $E_F$; black- (red-) line: spin-up (spin-down) channel.

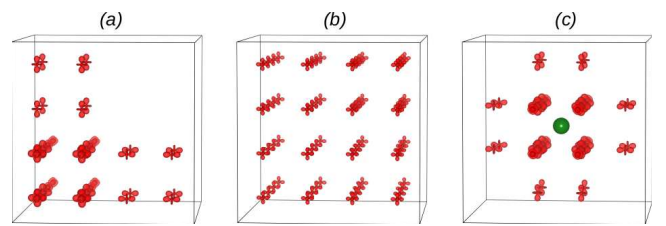

FIG. 15. Isosurfaces of electron density ($0.004$ Bohr$^{-1}$) of the additional carriers in the $2 \times 2 \times 2$ supercell, , as calculated using B1-WC, for: *(a)* doped with one additional electron in its FM ground state; *(b)* doped with one additional electron in the NM phase and *(c)* explicitly doped $Fe_2TiSn_{Sb}$ at $x = 1/32$, the Sb atom is displayed in green.

FIG. 11. Electron density of the additional electrons as calculated from B1-WC, at x = 1/32, computed on the (001) planes at different distances from the origin, corresponding respectively to Fe, $X/Y$ and Fe atomic planes. The later is the closest Fe (grey spots) plane to the impurity $A$ (green spot). At the bottom, half-cut of the cell is shown with the atoms indicated in the bottom-left legend highlighting with colors (red, blue and green) the planes where the charge density has been projected.

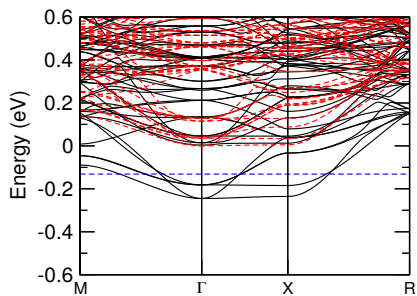

FIG. 12. Spin resolved band structure of $Fe_2TiSn_{As}$ ($x = 1/32$).

### Artificial doping within GGA+U, origin of magnetization

In Fig. 16, we show the spin-polarized DOS at $E_F$ and the cell magnetization as a function of the electron doping concentration, injected in pristine hosts (without atomic substitution), for the whole series of compounds. As in the main text, we start from the doped $Fe_2VAl$ ((c) panel) which shows a large distance between the two $e_g$ bands at the X point (larger than the case shown in Fig. 4(a), main text). Here, the additional electrons populate the V $e_g$ levels at the CBM and, consequently, no spin-splitting is induced up to $3.0 \times 10^{21}$ cm$^{-3}$, perfectly in agreement with B1-WC explicit doping, and as expected from fully delocalized states. Increasing the amount of doping more than $5.7 \times 10^{21}$ cm$^{-3}$ the system results unstable (large cell pressures) and its charge density is difficult to converge. In the Fe$_2$NbGa (Fig. 16(e)) no spin-polarization manifests while the Nb $e_g$ band is populated and Fe one remains empty (analogously to Fig. 4(a) and (b), main text). The magnetic transition appears when electrons are allocated on the Fe $e_g$ band with high local magnetic moments with interaction mediated by the itinerant electrons on the Nb site. Here, in fact, the validity of the Stoner condition ($St > 1$, see Eq. (5) and main text) is reached at $2.1 \times 10^{21}$ cm$^{-3}$ (highlighted with orange dashed line). Fe$_2$TaGa (Fig. 16(d)), analogously to Fe$_2$VAl and Fe$_2$NbGa, shows the Ta $e_g$ state at the CBM. However, differently to those cases, the distance between the two bands is lower resulting in a stronger hybridization (as Fig. 4(c) main text). This particular band relative position has very interesting consequences: the system goes across three main regimes induced by doping: at low concentrations, we start to populate the Ta $e_g$ level and at about $1.7 \times 10^{20}$ cm$^{-3}$ (maroon dot-dashed line) an exchange splitting is induced on Fe $e_g$ states although the splitting on the occupied Ta orbitals remains negligible. This behaviour confirms the interplay between orbital hybridization and the exchange interaction: in this region (maroon background, Fig. 16(d)) the direct exchange dominates ($St < 1$): there are not enough delocalized states to favour the itinerant electron with respect to the direct exchange. For this reason, when the Fe $e_g$ start to be populated (at critical concentration of about $1.9 \times 10^{20}$ cm$^{-3}$, starting point of the cyan background region in Fig. 16(d)), its strongly localized nature invalidates even more the itinerant picture ($St < 1$). Increasing the doping density at about $4.2 \times 10^{20}$ cm$^{-3}$ (orange dashed line), an sufficient number delocalized electrons makes the itinerant magnetism dominate and the Stoner condition $St > 1$ fulfilled. For the last two compounds, we tuned the $U_{Ti}$ to obtain a similar arrangement of the band structure as obtained with B1-WC, keeping the $U_{Fe}$ at the self-consistently estimated value. Fe$_2$TiSn and Fe$_2$TiSi (Fig. 16(a), same as Fig. 4(d) of main text, and (b)), having the Fe $e_g$ band at the CBM, show an induced spin-polarization and a consequent magnetization immediately (similarly to Fig. 4(d), main text) and in both cases their DOS confirm the acquired half-metallic character. The highly localized nature of the Fe $e_g$ orbitals and the strong exchange interaction among belonging carriers drive this behaviour. In the low doping region, moreover, the prevalent nature of the exchange interaction, still due to the localized nature of these states, is direct ($St < 1$). However, at about $3.8 \times 10^{20}$ cm$^{-3}$ for Fe$_2$TiSi and $3.6 \times 10^{20}$ cm$^{-3}$ for Fe$_2$TiSn, the itinerant exchange starts to dominate ($St > 1$) due to the presence of a sufficient number of free-like electrons.

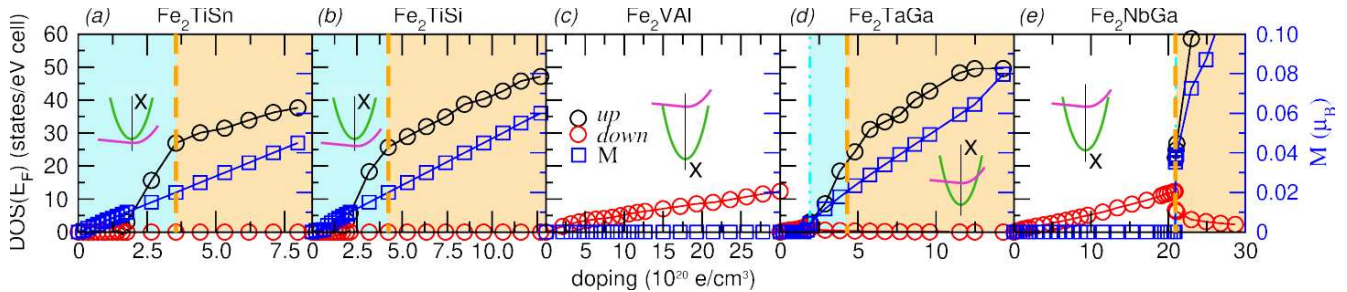

FIG. 16. GGA+$U$ results: spin-projected DOS at $E_F$ and cell magnetization versus doping for the five compounds: black (red) open circles refer to up (down) spin channels (left axis), open squares (blue) to the cell magnetization (blue right axis). Cyan double-dot-dashed line: critical carrier concentration needed to begin to populate the Fe $e_g$ state; orange dashed line: critical doping for which the Stoner criterion is verified ($St > 1$).

| Fe$_2YZ$ | $\Delta$E$_{ex}$ (eV) | g$^{NM}$(E$_F$)(st./eV) | $St$ |
|---|---|---|---|
| Fe$_2$TiSn | 0.200 | 35.31 | 7.1 |
| Fe$_2$TiSi | 0.154 | 37.84 | 5.8 |
| Fe$_2$VAl | 0.000 | 8.07 | 0.0 |
| Fe$_2$TaGa | 0.097 | 25.14 | 2.4 |
| Fe$_2$NbGa | 0.060 | 13.76 | 0.8 |

TABLE III. Evaluation of the Stoner criterion using Eq. (5): exchange splitting $\Delta$E$_{ex}$ and the value of the NM phase DOS at $E_F$ g$^{NM}$(E$_F$) calculated using B1-WC.

### Thermoelectric properties

In a rigid band picture, doping with donors (acceptors) only shift the chemical potential $\mu$ from its initial position the gap to bring it closer to the conduction (valence) band. In the case of Fe$_2YZ$ compounds, given the nature of the band edge (Fe 3$d$), substitution at the $Z$ site is expected to bring $\mu$ inside the conduction band, close to the band edge. This effect is highlighted in Fig. 17 for the specific case of Fe$_2$TiSn$_{Sb}$ at 300 K: the Seebeck coefficient $S$, electrical conductivity $\sigma$, the correspond-ing power factor $S^2\sigma$ as well as the electronic DOS are compared among: the pristine eigen-energies ($x = 0$), the NM phase and the FM phase concentration 1/16. For the explicitly doped case, $E_F$ lies inside the conduction band, as discussed in the previous section. The calculated values for the transport coefficients are very close to the rigid band predictions in the NM constrained phase; however, for the FM phase (ground-state), the in-gap states reduce the number of carriers (in the $k_B T$ interval around $E_F$) contributing to $S$ and $\sigma$, with $\mu$ near the conduction states, resulting in a decrement of the PF peak in the $n$-type region. In contrast, $S$ and $\sigma$ related a hypotetic $\mu$ lying in the $p$-type region, remain, in both cases, unaffected. We also show the spectral PF of the doped FM phase at 600 K with a corrected relaxation time $\tau$ to account for phonon scattering at high temperature [1], showing that the peak of power factor remains roughly around ~3 - 4 mW/K$^2$m at higher temperatures.

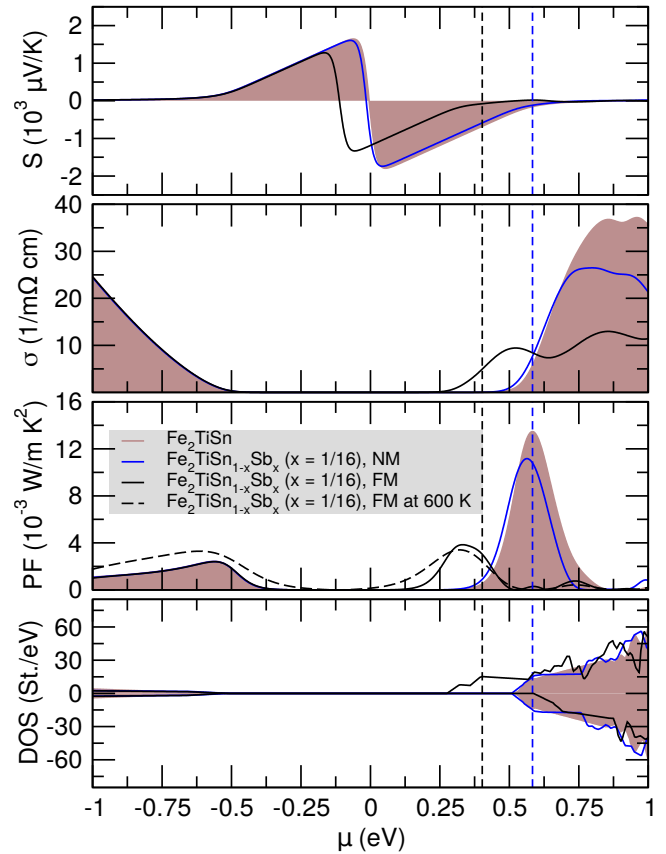

FIG. 17. Room-temperature Seebeck coefficient, electrical conductivity, power factor with respect to chemical potential $\mu$ of Fe$_2$TiSn$_{Sb}$ for $x = 0$ (brown background) and 1/16, NM (blue) and FM (black) magnetic phases. The associated electronic DOS are also given (with positive and negative value for spin-up and spin-down electron respectively. The zero energy is set as $E_F$ of the pristine material, the vertical dashed lines are the Fermi level $E_F$ for the doped phases. The transport properties have been computed using the B1-WC band structures.